\documentclass[preprint2,12pt]{aastex6}

\newcommand{\tco}{$^{13}$CO}
\newcommand{\co}{$^{12}$CO}
\newcommand{\ceo}{C$^{18}$O}
\newcommand{\tcoone}{$^{13}$CO~$J$=1$-$0}
\newcommand{\coone}{$^{12}$CO~$J$=1$-$0}
\newcommand{\ceoone}{C$^{18}$O~$J$=1$-$0}

\newcommand{\cothree}{$^{12}$CO~$J$=3$-$2}

\newcommand{\cosix}{$^{12}$CO~$J$=6$-$5}

\newcommand{\lfir}{$L_{\rm{FIR}}$}
\newcommand{\lsol}{$L_{\odot}$}
\newcommand{\msol}{$M_{\odot}$}

\newcommand{\kms}{km~s$^{-1}$}
\newcommand{\tkin}{$T_{\rm{kin}}$} 
\newcommand{\nhtwo}{$n_{\rm{H_{2}}}$}

\newcommand{\xco}{[$^{12}$CO]/[$^{13}$CO]}
\newcommand{\xceo}{[$^{13}$CO]/[C$^{18}$O]}

\shorttitle{Isotopes in IRAS~13120-5453}
\shortauthors{Sliwa et al.}

\begin{document}


\title{Extreme CO Isotopic Abundances in the ULIRG IRAS~13120-5453: An Extremely Young Starburst or Top-Heavy Initial Mass Function}

\author{Kazimierz Sliwa\altaffilmark{1},  
	Christine D. Wilson\altaffilmark{2},
	Susanne Aalto\altaffilmark{3},
	George C. Privon\altaffilmark{4,5}
	}
\altaffiltext{1}{MPI for Astronomy, K{\"o}nigstuhl 17, D-69117 Heidelberg, Germany; sliwa@mpia-hd.mpg.de}
\altaffiltext{2}{Department of Physics and Astronomy, McMaster University, Hamilton, ON L8S 4M1 Canada}
\altaffiltext{3}{Department of Earth and Space Sciences, Chalmers University of Technology, Onsala Space Observatory, SE-439 94 Onsala, Sweden}
\altaffiltext{4}{Instituto de Astrof\'sica, Facultad de Física, Pontificia Universidad Cat\'olica de Chile, Casilla 306, Santiago 22, Chile}
\altaffiltext{5}{Departamento de Astronom\'ia, Universidad de Concepci\'on, Casilla 160-C, Concepci\'on, Chile}




%


\begin{abstract}
We present ALMA \co\ (J=1-0, 3-2 and 6-5), \tco\ (J=1-0) and \ceo\ (J=1-0) observations of the local Ultra Luminous Infrared Galaxy, IRAS~13120-5453. The morphologies of the three isotopic species differ, where \tco\ shows a hole in emission towards the center. We measure integrated brightness temperature line ratios of \co/\tco\ $\geq$ 60 (exceeding 200) and \tco/\ceo\ $\leq$ 1 in the central region. Assuming optical thin emission, \ceo\ is more abundant than \tco\ in several regions.  The abundances within the central 500 pc are consistent with enrichment of the ISM via a young starburst ($<$7Myr), a top-heavy initial mass function or a combination of both.
\end{abstract}





\section{Introduction} 

Isotopic abundances in the interstellar medium (ISM) can be used as a tracer of stellar nucleosynthesis. The $^{12}$C/$^{13}$C isotope ratio is an important tracer of the relative degree of primary versus secondary processing in stars. The $^{12}$C atom is a primary species produced in intermediate and high-mass stars \citep[e.g][]{Prantzos1996}. Massive stars are also responsible for the majority of $^{16}$O and $^{18}$O. The $^{13}$C atom is an intermediary species that is transformed into $^{14}$N. In the red giant phase of low/intermediate mass stars, $^{13}$C is lifted to the envelope via convection \citep{Wilson1992} and eventually released into the ISM. Massive stars are short-lived and will start to enrich the ISM in $^{12}$C in $\sim$10$^{6}$ years while $^{13}$C enrichment needs $\sim$10$^{9}$ years \citep[e.g.][]{Vigroux1976}.  


Ultra/Luminous Infrared Galaxies (U/LIRGs) are extreme starbursts offering great laboratories to study high-mass star formation. It has long been observed that \tco\ emission is unusually weak relative to \co\ \citep[$\sim$ 20-40;][]{Aalto1991,Casoli1992} compared to that from normal disk galaxies \citep[$\sim$ 10; e.g.][]{Paglione2001}. Pioneering work by \cite{Casoli1992} and \cite{Henkel1993} presented several scenarios to explain this unusual emission ratio such as optical depth effects, abundance variations via some mechanism such as photo-dissociation, inflowing low metallicity gas or enrichment of the ISM and a two-phase molecular medium consisting of a diffuse envelope where \co\ can better self-shield than \tco. Recent radiative transfer modeling of the molecular gas in several U/LIRGs is consistent with high \xco\footnote{Square brackets denote an abundance ratio while all other ratios are integrated brightness temperature ratios} abundance ratios \citep[$>$90;][]{Sliwa2013,Sliwa2014,Henkel2014,Papadopoulos2014,Tunnard2015b}. 

Recent work on the [$^{16}$O]/[$^{18}$O] abundance in U/LIRGs using $Herschel$ H$_{2}$O and OH observations have shown varying values from $\leq$30 for Mrk231 \cite{GA2010}, around 50-150 for Arp220 \citep{GA2014} and Zw 049.057 \citep{Falstad2015} and $\geq$500 for Arp 299 \citep{Falstad2017} and NGC 4418 \citep{GA2014}. \cite{Koenig2016a} used Atacama Large Millimeter/submillimeter Array (ALMA) CO data to show that the [$^{16}$O]/[$^{18}$O] $\geq$ 900 for NGC~1614.  The high [$^{16}$O]/[$^{18}$O] abundance ratios are believed to be due to inflowing gas and the low values to stellar processing.

\objectname{IRAS~13120-5453} (dubbed the ``Yo-Yo") is a nearby ULIRG (D$_{\rm{L}}$~=~144~Mpc) with a far-infrared luminosity \citep[\lfir~=~1.5~$\times$~10$^{12}$~\lsol;][]{Sanders2003} similar to that of Arp~220. The system has been classified as a post-merger \citep{Haan2011}. X-ray emission is consistent with a Compton-thick active galactic nucleus \citep[AGN;][]{Iwasawa2011} and contributes $\sim$ 18$\%$ to the infrared luminosity \citep{Sturm2011}. The $Herschel$ Fourier Transform Spectrometer (FTS) observed multiple high-$J$ CO lines as well as CI, H$_{2}$O, NII, OH and more  \citep{Rosenberg2015,Mashian2015,Pearson2016,Privon2017}.  \cite{Privon2017} shows that the HCN/HCO$^{+}$ line ratio observed with ALMA suggests an increased HCN abundance via turbulent heating.

In this Letter, we present new ALMA Cycle~2 observations of IRAS 13120-5453 where we have detected three \co\ transitions, \tco\ and \ceoone. The morphology of the three species differs and offers insight into the mechanism that may be controlling the \xco\ ratio. We show that massive stars have enriched the ISM in $^{12}$C and $^{18}$O and drive the observed line ratios in IRAS~13120-5453.

\section{Observations and Line Ratios}  

\begin{figure*}[htbp] 
\centering
\gridline{\fig{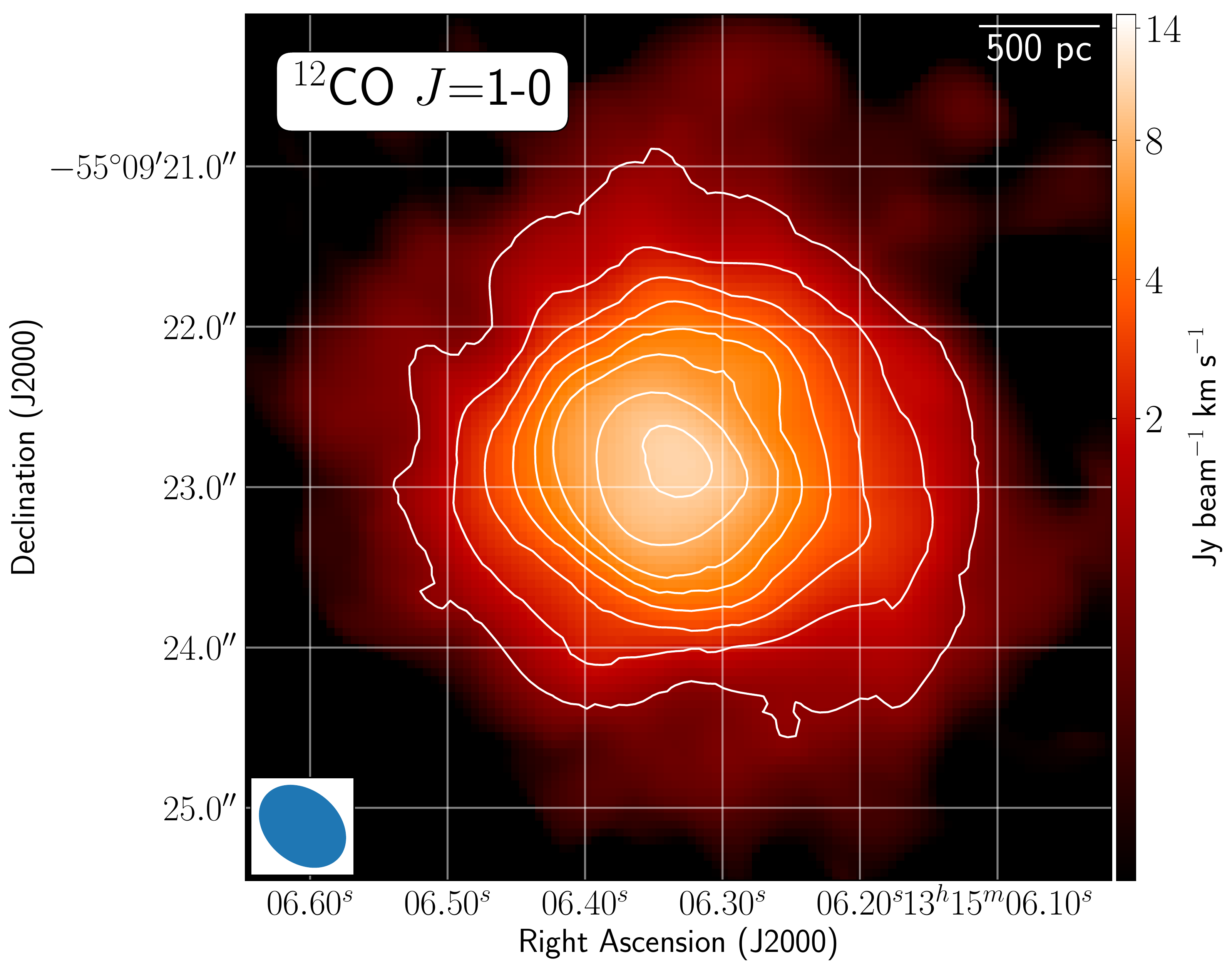}{0.45\textwidth}{(a)} \fig{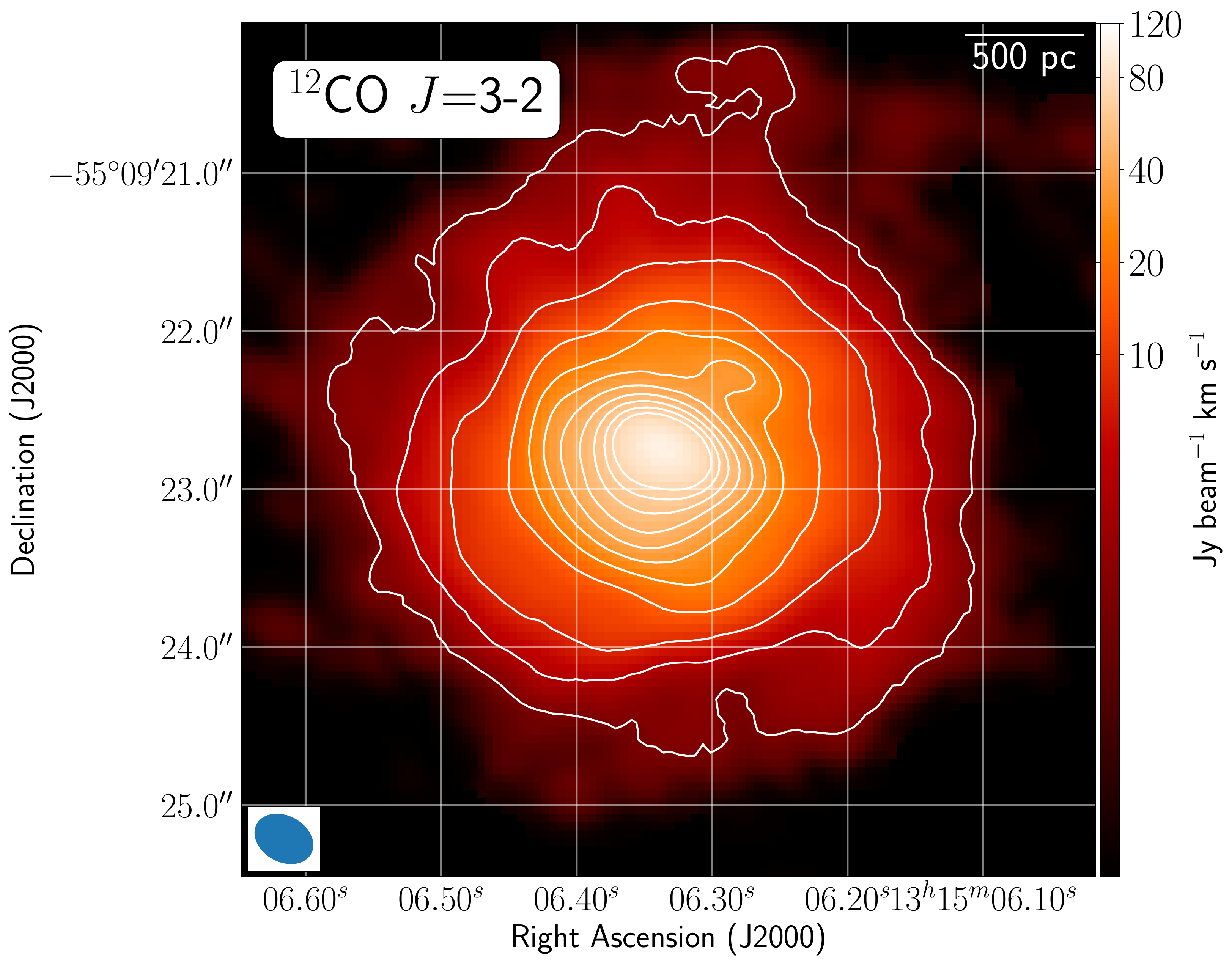}{0.45\textwidth}{(b)} }
\gridline{\fig{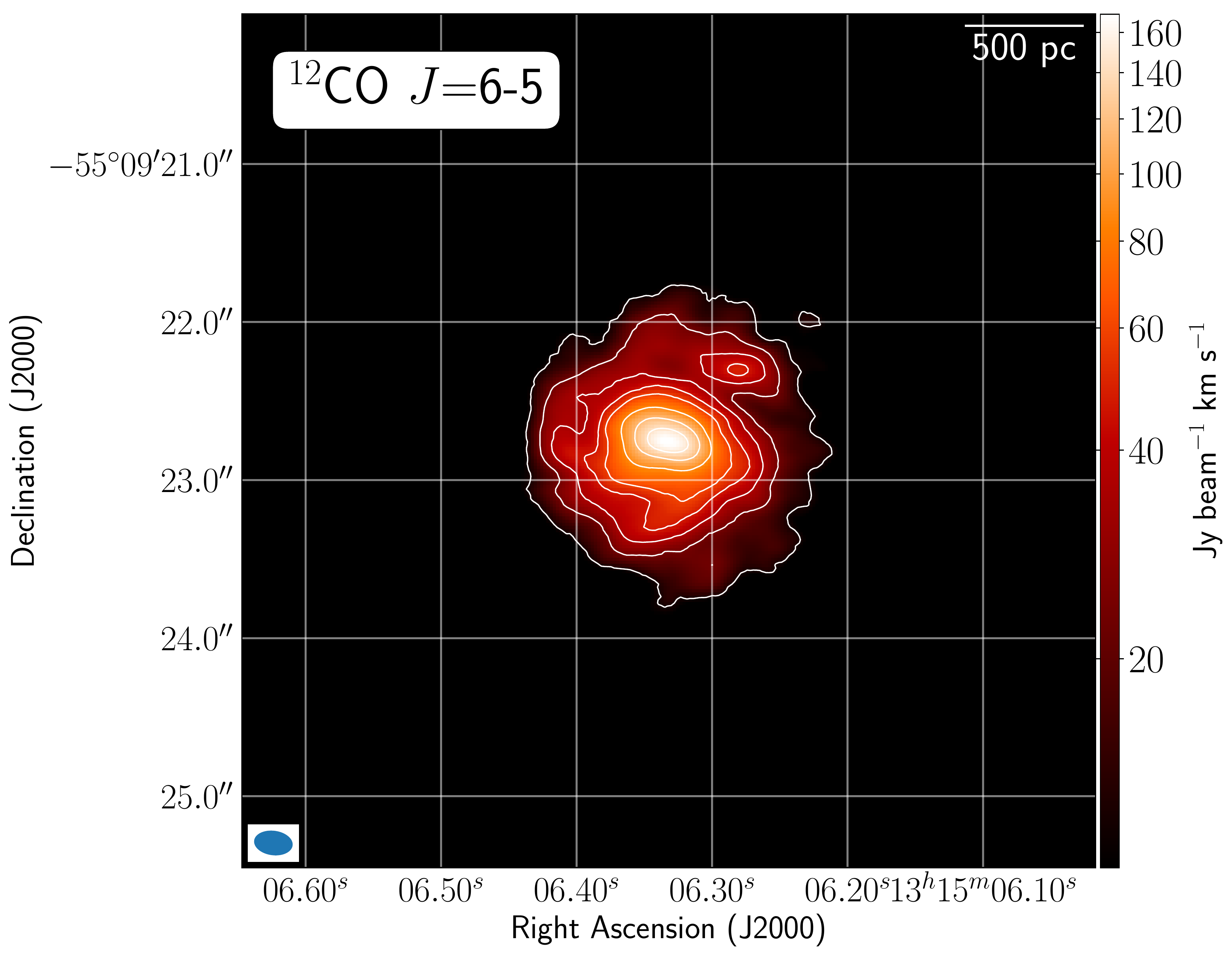}{0.45\textwidth}{(c)} \fig{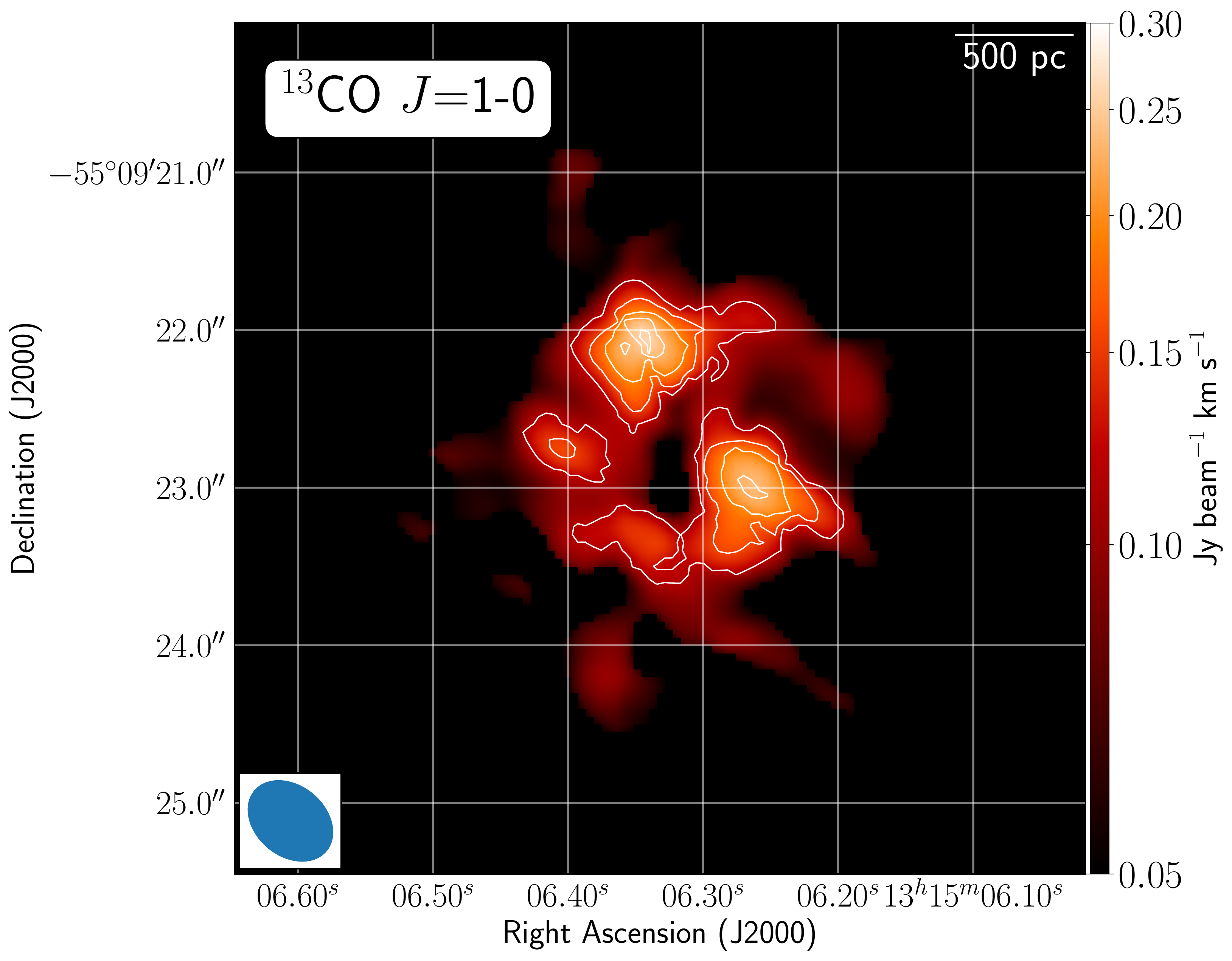}{0.45\textwidth}{(d)} }
\gridline{\fig{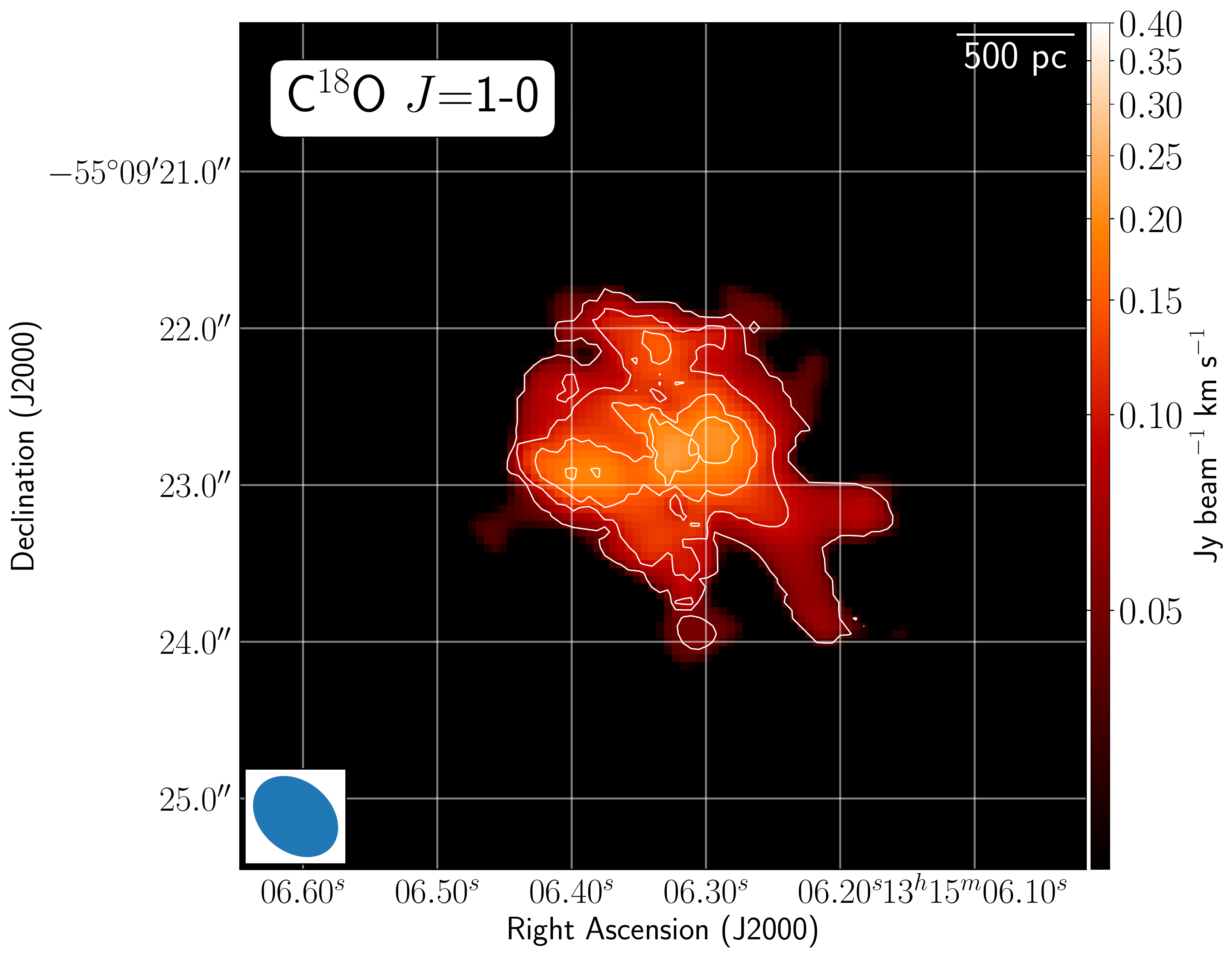}{0.45\textwidth}{(e)}  }
\caption{Integrated intensity maps for IRAS~13120-5453: (a) \coone, (b) \cothree, (c) \cosix, (d) \tcoone\ and (e) \ceoone. Ellipse in the bottom left corner represents the synthesized beam.}
\label{fig:linemaps}
\end{figure*}

ALMA was used to observe IRAS 13120-5453 in Cycle 2 using Bands 3, 7 and 9 (Table \ref{tab:fluxdata}). We calibrated all datasets manually in CASA v4.5.3 \citep{McMullin2007} using standard calibration steps. 
We implemented two iterations of phase-only self calibration on the \co\ datasets, which did not significantly alter the morphology of IRAS 13120-5453.  We CLEAN the datacubes using a Briggs robust weighting of 0.5 down to 1$\sigma$ level with channels widths of 20 or 35 \kms. Integrated intensity maps were created using the CPROPs \citep{Rosolowsky2006} masking routine\footnote{The routine finds pixels greater than 3$\sigma$ in 2 channels and then includes emission down to some $\sigma$ level around the pixel. This method is excellent at excluding spurious noise pixels} and only channels that contained masked emission were included down to 1.5$\sigma$. All maps were primary beam corrected (Figure \ref{fig:linemaps}). 

\begin{deluxetable*}{cccccc} 
\tablewidth{0pt}
\tabletypesize{\scriptsize}
\tablecaption{Observational Data \label{tab:fluxdata}}
\tablehead{
\colhead{Parameter\tablenotemark{a}} & \colhead{\coone} & \colhead{\cothree} & \colhead{\cosix} & \colhead{\tcoone} & \colhead{\ceoone} }
\startdata
Obs Data & 2015 July 03	& 2015 June 07	&2015 June 09	&2015 July 22&2015 July 22				\\
Calibrators &J1107-4449	&J1427-4206	&J1256-0547		& J1107-4449	&J1107-4449	\\
 		&J1551-1755	&Titan		&J1427-421		&J1427-421	&J1427-421	\\
 		&Titan		&J1329-5608 	&J1329-5608		&J1329-5608	&J1329-5608	\\
		&J1329-5608	&J1315-5334 	&J1427-4206	 	& 			&  \\
Integration Time (s) & 652 & 867 & 1567 & 3117 & 3117 \\
Median PWV (mm) & 2.4 & 0.63 & 0.37 & 3.2 & 3.2 \\
Median T$_{\rm{sys}}$ (K) & 64 & 135 & 984 & 82 & 82 \\
Flux\tablenotemark{b} (Jy \kms) &126 ($\pm$2) [$\pm$ 13]	&1265 ($\pm$ 15) [$\pm$ 130] 	&2460 ($\pm$	25) [$\pm$ 370]	& 2.21 ($\pm$0.05) [$\pm$0.2] & 2.1 ($\pm$0.05) [$\pm$0.2] \\
rms (mJy beam$^{-1}$)&1.5 (20~\kms)	&1.1 (20~\kms) &16 (20~\kms)&   0.3 (35~\kms) & 0.3	 (35~\kms)\\
Resolution (arcsec)&	 0.58 $\times$ 0.35	&0.39 $\times$ 0.29& 0.25$\times$ 0.16	& 0.55 $\times$ 0.41 &0.55 $\times$ 0.41	\\
\enddata
\tablenotetext{a}{Other line present in the data will be discussed in forthcoming paper (Sliwa et al in prep.) }
\tablenotetext{b}{Uncertainties in curved and square brackets denote measurement and calibration uncertainties, respectively.  }
\end{deluxetable*}

\begin{figure*}[htbp] 
\centering
\fig{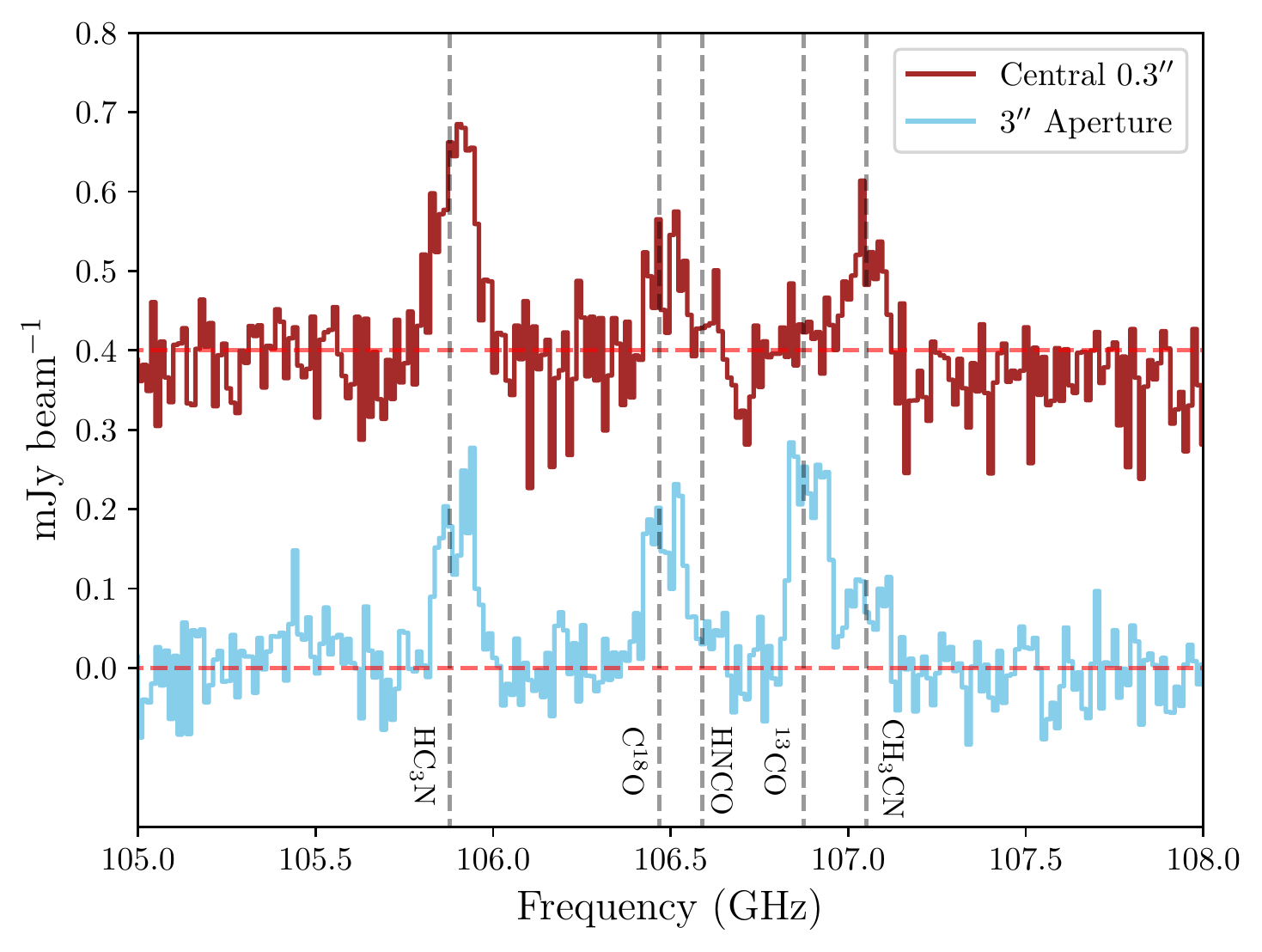}{1\textwidth}{} 
\caption{Averaged spectra for (top) the central 0.3\arcsec\ diameter aperture scaled down by a factor of 6 and shifted up by 0.4 mJy~beam$^{-1}$ and (bottom) over a 3\arcsec\ diameter aperture. Identified spectral lines are marked with vertical dashed lines (z=0.03112). For HNCO, we mark the 5$_{0,5}$-4$_{0,4}$ transition.}
\label{fig:spectra}
\end{figure*}

All three \co\ transitions have similar morphologies with a single nucleus. 

The \tcoone\ emission is more interesting with no emission above 1.5$\sigma$ near the central nucleus ($\sim$ 0.85\arcsec\ $\times$ 0.3\arcsec\ = 590 pc $\times$ 210 pc) and two relatively strong emission regions outside the nucleus. The region lacking emission is within the starburst region (0.5~kpc) measured by \cite{Privon2017}. \textit{Interestingly, the C$^{18}$O emission is relatively strong where there is no $^{13}$CO emission}.  \textit{In the central 0.3\arcsec\ spectrum (Figure \ref{fig:spectra}), it is evident that C$^{18}$O is stronger than $^{13}$CO.} We note that \ceo\ may be partially contaminated by HNCO (5$_{0,5}$ - 4$_{0,4}$) and several higher-energy transitions that lie on top of the \ceo\ line. Along a line of sight, the maximum contamination is likely 30$\%$ determined from the peak of the HNCO (5$_{0,5}$ - 4$_{0,4}$) transition; however, we do not make any corrections since this is an upper limit to the contamination and may only contribute to part of the \ceo\ line profile. 

Selective UV photodissociation of the rare CO isotopologues should affect both \tco\ and \ceo, with the nominally rarer \ceo\ affected the most. Thus it could not produce the observed relative line intensity ratio variations between them, let alone boost the \ceo\ abundance to be higher than \tco\ in the inner 500~pc of IRAS 13120-5453. If optical depth effects were causing the ring, \ceo\ should also be observed in a ring since both \tco\ and \ceo\ are assumed to be optically thin. The Band 3 observations, observed 19 days apart, have similar $uv$-coverage adding confidence to the observed differences in morphology and corresponding intensities between the lines. The three \co\ maps (Figure \ref{fig:linemaps}) also show that as we go to higher-resolution (J=1-0 $\rightarrow$ J=6-5) we do not see a ring in \co. 

\begin{figure}[htbp] 
\gridline{\fig{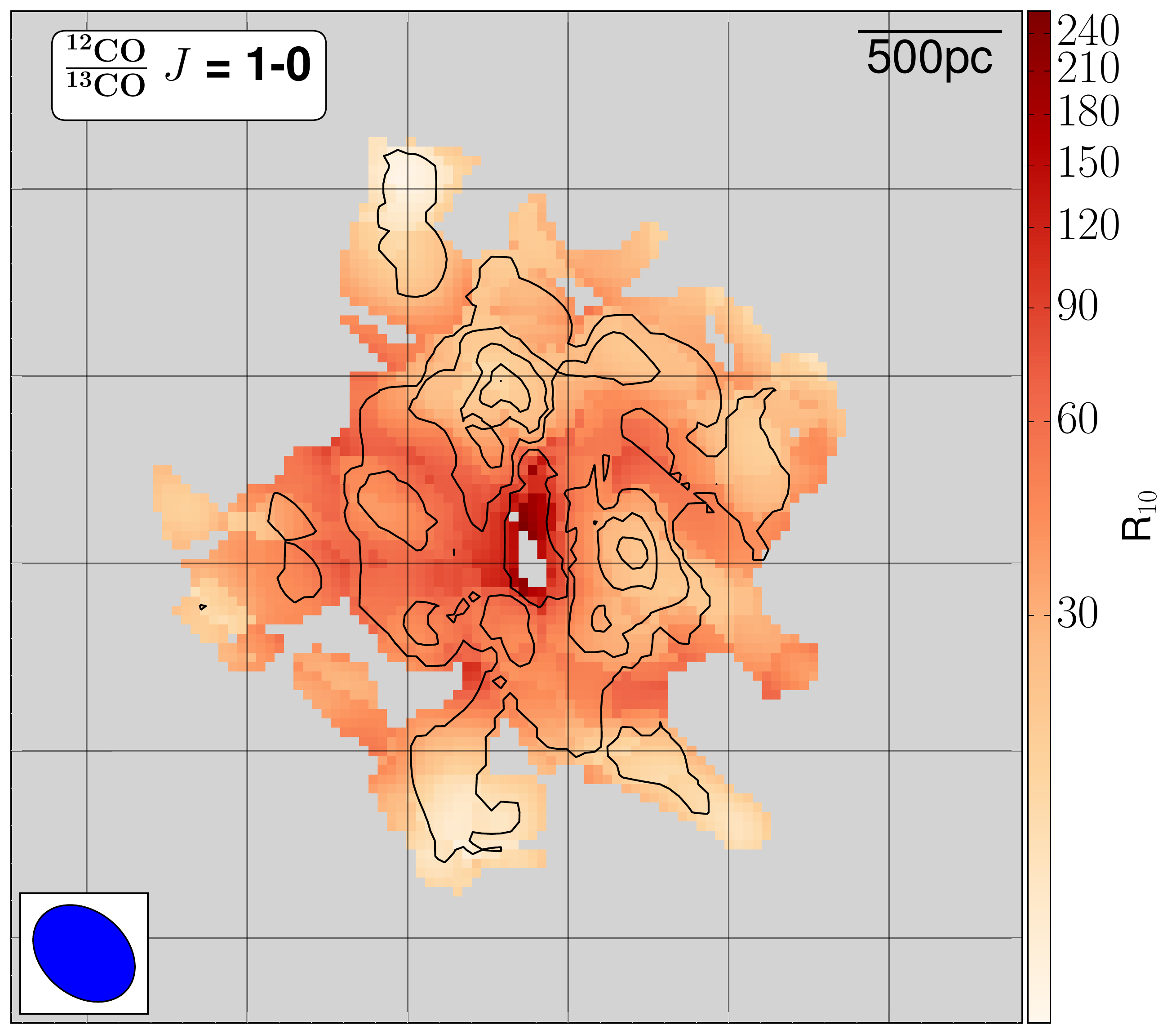}{0.4\textwidth}{(a)}}
 \gridline{\fig{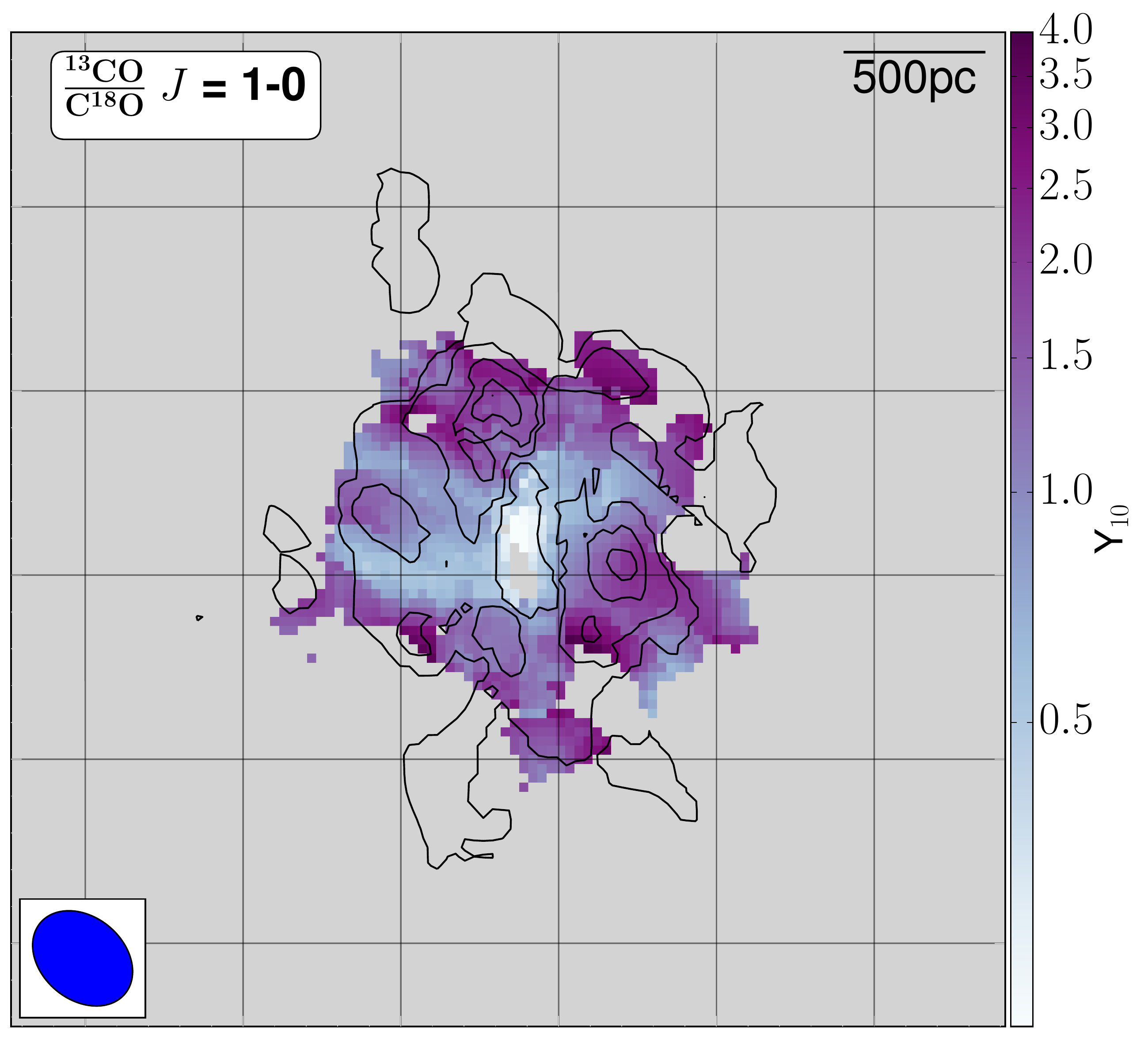}{0.4\textwidth}{(b)} } 
\gridline{\fig{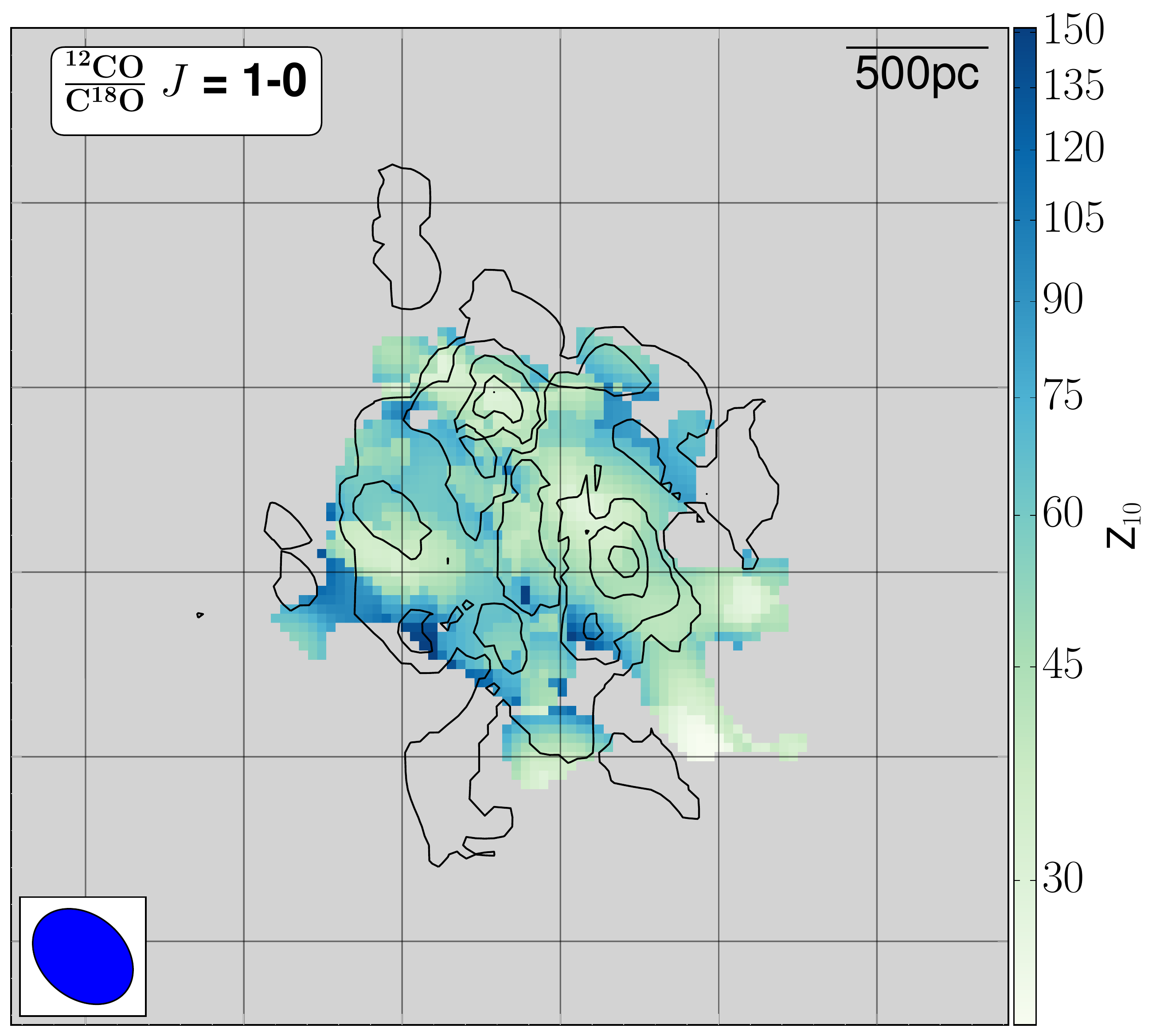}{0.4\textwidth}{(c)} } 
\caption{Line brightness temperature ratio maps for (a) R$_{10}$, (b) Y$_{10}$ and (c) Z$_{10}$. Each map has an angular resolution of 0.6\arcsec $\times$ 0.45\arcsec. Black contours are of \tcoone\ added as a guide.}
\label{fig:ratios}
\end{figure}

Integrated brightness temperature (I = $\int$T$_{B}$dV) line ratio maps can offer some insight into the conditions of the molecular gas. We create the following line ratios maps:

\noindent R$_{10}$ = $\frac{I^{^{12}CO(1-0)}}{I^{^{13}CO(1-0)}}$ \\
Y$_{10}$ = $\frac{I^{^{13}CO(1-0)}}{I^{C^{18}O(1-0)}}$  \\
Z$_{10}$ = $\frac{I^{^{12}CO(1-0)}}{I^{C^{18}O(1-0)}}$  \\
We match the \co, \tco\ and \ceoone\ maps to an angular resolution of 0.60\arcsec\ $\times$ 0.45\arcsec. We then create line ratio maps by cutting emission below 2$\sigma$ in each map and converting the units from Jy~beam$^{-1}$ \kms\ to K(T$_{B}$) \kms\ (Figure \ref{fig:ratios}). 

The R$_{10}$ line ratio shows a wide range of values from $\sim$10 to over 250. While values of around 30 are common for local U/LIRGs \citep[e.g][]{Sliwa2012,Sliwa2013,Sliwa2014}, values exceeding 100 have never been observed before in local ULIRGs. The Y$_{10}$ line ratio ranges from 0.2 to over 4. Values below 1 are rare in extragalactic systems where normal disk galaxies have an average  Y$_{10}$ value of $\sim$ 6 and normal starbursts show Y$_{10}$ values of $\sim$ 3 \citep{JD2017}. Arp~220 \citep{Greve2009,Matsushita2009} and the high-z ULIRG SMM~J2135-0102 \citep{Danielson2013} show a Y$_{10}$ ratio of 1 while the LIRG merger remnant, NGC~2623, shows a Y$_{10}$ ratio of $\sim$ 1.8 (Sliwa et al. in preparation). 
The Z$_{10}$ values range from $\sim$ 20-140 and are similar to those of Mrk~231 \citep{GA2010}, Arp~220 \citep{GA2014} and Zw 049.057 \citep{Falstad2015}.  

\section{Extreme Isotopic Abundances}
We argue that IRAS 13120-5453 has an extreme isotopic abundance ratio when compared to normal star-forming galaxies. The brightness temperature line ratio of species A and B can be expressed as
\begin{equation}
R = \frac{I^{A}}{I^{B}} = \frac{T_{\rm{ex}}^{A}}{T_{\rm{ex}}^{B}}\frac{(1 - e^{-\tau_{A}})}{(1 - e^{-\tau_{B}})}
\end{equation}
where T$_{\rm{ex}}$ is the excitation temperature and $\tau$ is the optical depth of the particular species and transition. For simplicity, we assume local thermal equilibrium (LTE; i.e. T$_{\rm{ex}}^{A}$ = T$_{\rm{ex}}^{B}$ = T$_{\rm{kin}}$). (Significant radiative trapping for the abundant species can give T$_{\rm{ex}}^{A}$ =  T$_{\rm{kin}}$ while T$_{\rm{ex}}^{B}$ $<$ T$_{\rm{kin}}$.)    If species B is optically thin and species A is optically thick, then
\begin{equation}
R \sim \frac{1}{\tau_{B}} \sim \frac{[A]}{[B]}\frac{1}{\tau_{A}}.
\end{equation}
If \co\ is optically thick, the line ratios R$_{10}$ and Z$_{10}$ are lower limits to the relative abundance of \co\ to \tco\ and \ceo\ since the observed line ratio is attenuated by the optical depth of \co. Since R$_{10}$ and Z$_{10}$ $>>$ 1, both \tco\ and \ceo\ must be optically thin. Thus the Y$_{10}$ line ratio of $<$ 1  implies that {\it \ceo\ is more abundant that \tco\ in the central region;} this would still be true even if the HNCO contamination were to reach 50\%.

\section{Root of the Extreme Abundances}

\textit{Photo-dissociation:} Since \ceo\ is bright relative to \tco, we can rule out selective photo-dissociation as the dominant mechanism since both \ceo\ and \tco\ would be destroyed by UV radiation. 

\textit{Fractionation:} The most important carbon isotope exchange is 
\begin{equation}
\rm{^{13}C^{+}} + \rm{^{12}CO} \rightleftharpoons \rm{^{13}CO} + \rm{^{12}C^{+}} + \Delta E
\end{equation}
 \citep{Watson1976} where the forward reaction dominates in cold environments ($<$ 30 K) favouring the formation of \tco. In hot environments, both directions have equal probability \citep{Roueff2015}.  
We use the non-LTE code RADEX \citep{vanderTak2007} and a Bayesian likelihood code \citep{Kamenetzky2012} to constrain the molecular gas physical conditions within the $\sim$400pc central region using only the \co\ observations. We fit the \coone\ line with a Gaussian profile of FWHM=375 \kms. The most probable solution is warm, dense molecular gas with a \tkin\ = 130$^{+400}_{-77}$ K , log(\nhtwo) = 4.2$^{+2.4}_{-0.0}$ cm$^{-3}$ and log(M(H$_{2}$)/\msol) = 7.8$^{+1.0}_{-0.1}$ . Since the molecular gas is not cold enough for the forward reaction to dominate, we can rule out fractionation as a possible mechanism affecting the abundance. We also note that \ceo\ does not undergo fractionation and should reflect stellar processing \citep{Langer1984}.

\textit{Infalling Gas:} The merger process can drive a gas inflow towards the nuclear regions \citep[e.g.][]{Hopkins2006,Kewley2006,Ellison2008}. The Galaxy has an increasing radial gradient in the \xco\ abundance ratio ranging from 30 in the center to $>$100 at large radii \citep[e.g.][]{Milam2005}. \cite{JD2017} have shown that a trend with \xceo\ exists in disk galaxies as well with an average value of 6.0$\pm$0.9. Analyses of close galaxy pairs have shown that their metallicities are lower than similar field galaxies \citep{Kewley2006, Ellison2008}. \cite{Rupke2008} found that the dilution of the nuclear metallicity (Z) due to gas inflow is Z$_{final}$/Z$_{initial}$ $\sim$ 0.5; therefore, if we assume an initial \xco\ ratio of 30, we would expect a final ratio of $\sim$ 60. This would not be sufficient to explain the observed line ratios in the central regions, particularly the brighter \ceo\ emission.

\textit{Nucleosynthesis:} Enrichment of the ISM via massive stars is a likely mechanism. Massive stars are the dominant sources of $^{12}$C, $^{16}$O and $^{18}$O while $^{13}$C is predominately released from low/intermediate mass stars. Simulations show that the metallicity in the merger increases when the star-formation rate increases significantly, especially near the end of the merger process \citep{Torrey2012}. 

For nucleosynthesis enrichment to be plausible, the starburst must be young. With a normal initial mass function (IMF) such as the Kroupa IMF \citep[e.g.][]{Kroupa2001} within $\sim$6Myr, all stars $>$30\msol\ will have gone supernova ejecting material. Using the nucleosynthesis yield calculations for core-collapse supernovae (CCSNe) of \cite{Nomoto2006} and assuming a total star-forming molecular mass of 10$^{9}$\msol\ and an initial metallicity of Z=0.02 ($\sim$Z$_{\odot}$), the \xco\ value of the ejected material after 6Myr will be $\sim$575, while after 7Myr when all stars above 25\msol\ will have gone supernova, the \xco\ will be $\sim$60. 

If the starburst is older ($>$7Myr), an alternative solution is a top=heavy IMF.  \cite{Bartko2010} find a top-heavy IMF for the Galactic center of dN/dm $\propto$ m$^{-0.45 \pm 0.3}$. \cite{Habergham2010} also invoke a top-heavy IMF to explain the excess of CCSNe in interacting/mergers galaxies when compared to isolated galaxies. Assuming only stars of 10-130\msol\ will eject material vis CCSNe, a Kroupa IMF will produce a \xco\ abundance of 40 while a flat, top-heavy IMF ($\phi_{\rm{m}}$$\propto$m$^{0}$) will produce a \xco\ abundance of 270. While a flat IMF is an arbitrary choice, a top-heavy IMF of some variety is required if the starburst is older than $\sim$6Myr to explain our observed abundances in the central region. We also note that a combination of both a young-starburst and a top-heavy IMF is also plausible. Future work into the star formation history of IRAS 13120-5453 is required to clarify the starburst-age/IMF degeneracy.

\acknowledgments
We thank the referee for improving this manuscript.This paper uses the following ALMA data: 
ADS/JAO.ALMA$\#$2012.1.00306.S + ADS/JAO.ALMA$\#$2013.1.00379.S. ALMA is a partnership of ESO (representing its member states), NSF (USA) and NINS (Japan), together with NRC (Canada) and NSC and ASIAA (Taiwan) and KASI (Republic of Korea), in cooperation with the Republic of Chile. The JAO is operated by ESO, AUI/NRAO and NAOJ. The NRAO is a facility of the NSF operated under cooperative agreement by Associated Universities, Inc. G.C.P. was supported by a FONDECYT Postdoctoral Fellowship (No.\ 3150361).



 \facility{ALMA}

\clearpage

\end{document}